\documentclass[12pt]{article}
\pdfoutput=1
\usepackage{jheppub}
\usepackage{amsmath}
\usepackage{bm}

\usepackage{comment}

\newcommand{\tr}{\mathrm{tr}\,}

\newlength{\dummysp}
\usepackage[font=scriptsize]{caption}
\def\Z{{\mathbb Z}}

\def\tr{\,{\rm tr}\,}

\def\T{{\mathbb T}}

\def\beq{\begin{equation}}
\def\eeq{\end{equation}}

\title{Noninvertible anomalies in  $SU(N)\times U(1)$ gauge theories}
 \author[a]{Mohamed M. Anber,}\author[b]{Erich Poppitz} 
  
\affiliation[a]{Centre for Particle Theory, Department of Mathematical Sciences, Durham University, South Road, Durham DH1 3LE, UK}
\affiliation[b]{Department of Physics,   University of Toronto, 60 St George St., 
Toronto, ON M5S 1A7, Canada}
\emailAdd{mohamed.anber@durham.ac.uk}\emailAdd{poppitz@physics.utoronto.ca}    

\abstract{

{\flushleft{W}}e study $4$-dimensional $SU(N)\times U(1)$ gauge theories with a single massless Dirac fermion in the $2$-index symmetric/antisymmetric representations and show that they are endowed with a noninvertible $0$-form $\tilde {\mathbb Z}_{2(N\pm 2)}^{\chi}$ chiral symmetry along with a $1$-form $\mathbb Z_N^{(1)}$ center symmetry. By using the Hamiltonian formalism and putting the theory on a spatial three-torus $\mathbb T^3$, we construct the non-unitary gauge invariant operator corresponding to $\tilde {\mathbb Z}_{2(N\pm 2)}^{\chi}$ and find that it acts nontrivially in sectors of the Hilbert space characterized by selected magnetic fluxes. When we subject $\mathbb T^3$ to $\mathbb Z_N^{(1)}$ twists, for $N$ even, in selected magnetic flux sectors, the algebra of  $\tilde {\mathbb Z}_{2(N\pm 2)}^{\chi}$ and $\mathbb Z_N^{(1)}$ fails to commute by a $\mathbb Z_2$ phase. We interpret this noncommutativity as a mixed anomaly between the noninvertible and the $1$-form symmetries. The anomaly implies that all states in the torus Hilbert space with the selected magnetic fluxes exhibit a two-fold degeneracy for arbitrary $\mathbb T^3$ size. The degenerate states are labeled by discrete electric fluxes and are characterized by nonzero expectation values of condensates. In an Appendix, we also discuss how to construct the corresponding noninvertible defect via the ``half-space gauging'' of a discrete one-form magnetic symmetry.}

\begin{document}

\maketitle

%%%%%%%%%%%%%%%%%%
\section{Introduction}
%%%%%%%%%%%%%%%%%%%

Symmetries are the backbone of quantum field theory (QFT); the spectrum of local and extended operators are organized in symmetry representations. The modern way to define symmetry is via its action on topological surfaces. A $p$-form symmetry acts on $p$-dimenstional objects and is generated by operators supported on $p+1$-codimensional surfaces \cite{Gaiotto:2014kfa}. Traditionally, symmetries are connected to groups, and the operator that generates the symmetry is unitary. However, recent developments have highlighted the need to broaden the definition of symmetries to include actions generated by nonunitary operators.  These symmetries are noninvertible since the corresponding operators do not have an inverse.  Consequently, these operations do not form groups but can be comprehended as categories, offering a new perspective on the nature of symmetries.

Noninvertible symmetries were first identified and applied in $2$-dimensional QFT, see, e.g., \cite{Fuchs:2002cm, Komargodski:2020mxz}. The appreciation of the role of this new development in $4$-dimensional QFT resulted in an avalanche of works on this topic (a  non-comprehensive list is \cite{Nguyen:2021yld,Nguyen:2021naa,Choi:2021kmx,Wang:2021vki,Bhardwaj:2022yxj,
Choi:2022zal,Kaidi:2022uux,Choi:2022jqy,Cordova:2022ieu,Choi:2022rfe,Bartsch:2022mpm,Heckman:2022muc,Cordova:2022fhg,Karasik:2022kkq,GarciaEtxebarria:2022jky,Choi:2022fgx,Yokokura:2022alv,Bhardwaj:2022kot,Bhardwaj:2022maz,Bartsch:2022ytj,Heckman:2022xgu,Apte:2022xtu,Delcamp:2023kew,Kaidi:2023maf,Putrov:2023jqi,Dierigl:2023jdp}).

% \cite{Kaidi:2022uux,Cordova:2022ieu,Apte:2022xtu,GarciaEtxebarria:2022jky,Karasik:2022kkq,Choi:2022jqy,Choi:2021kmx,Yokokura:2022alv} 
One significant development in the field is the recognition that $4$-dimensional quantum electrodynamics (QED) possesses a noninvertible symmetry  \cite{Choi:2022jqy,Cordova:2022ieu}. The idea is that QED with a single Dirac fermion has a classical $U(1)_\chi$ chiral symmetry broken to the $\mathbb Z_2$ fermion number by the ABJ anomaly. The Noether current corresponding to $U(1)_\chi$ is not conserved as it receives a contribution from the anomaly. However, one may define a conserved chiral current by subtracting a Chern-Simons term that encodes the anomaly. Next, an operator corresponding to $U(1)_\chi$ is constructed by exponentiating the modified current and integrating it on a $3$-surface. This operator is not  gauge invariant. However, dressing it with a TQFT makes it gauge invariant. The resulting operator can be shown to generate a noninvertible symmetry for every rational value of the $U(1)_\chi$  parameter. The construction in \cite{Choi:2022jqy,Cordova:2022ieu} was further developed in \cite{Karasik:2022kkq,GarciaEtxebarria:2022jky} by showing that the definition of the noninvertible operator can be extended to any real parameter of $U(1)_\chi$ by coupling the theory to a scalar field living on the $3$-surface.

Comprehending this novel structure in gauge theories is important in pursuing a deeper understanding of QFT.  The present work examines $SU(N)\times U(1)$ gauge theory with a single massless Dirac fermion in a  representation $R$. The theory has a classical $U(1)_\chi$ symmetry broken by the ABJ anomaly in $SU(N)$- and $U(1)$-instanton backgrounds. Does this theory exhibit noninvertible symmetries, and if so, can they be utilized to establish exact nonperturbative statements? We demonstrate that the answer to this query is affirmative. 

 Unlike QED,  when our theory is put on a general manifold, the chiral symmetry is reduced to $\mathbb Z_{\scriptsize2\mbox{gcd}(T_R, d_R)}^\chi$, where $T_R$ and $d_R$ are the Dynkin index and dimension of $R$. Interestingly, we also show that the theory possesses a noninvertible $\tilde{\mathbb Z}_{2T_R}^\chi$ discrete chiral symmetry, wherein $\mathbb Z_{\scriptsize2\mbox{gcd}(T_R, d_R)}^\chi$ is an invertible part.\footnote{\label{fermionnumberfootnote}Throughout this paper, we use a tilde to distinguish noninvertible symmetries and operators. Note also that after gauging the vector $U(1)$, the $\mathbb Z_2$ part (fermion number) of $\tilde{\mathbb Z}_{2 T_R}^\chi$ is part of the gauge symmetry and not a global symmetry. With this in mind, we continue to denote the noninvertible symmetry by  $\tilde{\mathbb Z}_{2 T_R}^\chi$, as this subtlety does not affect our considerations of the anomaly and spectral degeneracy.  } We establish the noninvertibility through a sequential process starting from the nonconservation of Noether's current of the chiral symmetry. Employing the Hamiltonian formalism and putting the theory on a $3$-dimensional spatial torus $\mathbb{T}^3$, we construct a noninvertible symmetry operator. This setup provides a simple and explicit route to select the states on which the operator $\tilde{\mathbb Z}_{2T_R}^\chi$ acts nontrivially. 

$SU(N)\times U(1)$ QCD-like theories are naturally endowed with an electric $1$-form $\mathbb Z_N^{(1)}$ symmetry acting on the Wilson loops as well as a $1$-form $U(1)^{(1)}_m$ magnetic symmetry that characterizes sectors with definite magnetic fluxes. Then, it is natural to ask whether the theory exhibits a 't Hooft anomaly as we perform a $\tilde{\mathbb Z}_{2T_R}^\chi$ transformation in the background of $\mathbb Z_N^{(1)}$. To address this question, for definiteness\footnote{The conclusions of the work described can be generalized to fermions in higher representations.}
 we consider a theory with fermions in the $2$-index symmetric/anti-symmetric representation, which possesses a $\tilde{\mathbb Z}_{2(N\pm 2)}^\chi$ noninvertible symmetry.  We subject the theory to $\mathbb Z_N$ twists (2-form background fields for $\mathbb Z_N^{(1)}$) along the non-trivial cycles of  $\mathbb T^3$. In the presence of the  $\mathbb Z_N$ twists, the noninvertible symmetry projects onto sectors in Hilbert space with definite magnetic fluxes, easily identified in the Hamiltonian formalism. For $N$ even, the algebra of  $\tilde {\mathbb Z}_{2(N\pm 2)}^{\chi}$ and $\mathbb Z_N^{(1)}$ fails to commute by a $\mathbb Z_2$ phase inside these sectors, revealing a mixed anomaly between $\mathbb Z_N^{(1)}$ and $\tilde{\mathbb Z}_{2(N\pm2)}^\chi$ symmetries (anomalies involving noninvertible symmetries have previously been considered in $1+1$ dimensions, see \cite{Brunner:2014lua,Bhardwaj:2017xup,Chang:2018iay,Thorngren:2019iar,Zhang:2023wlu,Choi:2023xjw} and references therein).   This anomaly implies that the states in these special sectors must be $2$-fold degenerate\footnote{Similar to the exact finite-volume degeneracies due to invertible 0-form/1-form anomalies \cite{Cox:2021vsa}.}  on arbitrary size $\mathbb T^3$. Such degeneracies could be seen by examining the condensates in realistic lattice simulations.  

This paper is organized as follows. In Section \ref{gauge theories noninvertible symmetries}, we demonstrate the origin of the noninvertible symmetry by carefully examining $SU(N)\times U(1)$ QCD-like theories put on $\mathbb T^3$ with general flux backgrounds and build the noninvertible operator of $\tilde{\mathbb Z}_{2(N\pm 2)}^\chi$. Then, we show that this symmetry has a 't Hooft anomaly with the $\mathbb Z_N^{(1)}$ symmetry. In Section \ref{Hilbert space, magnetic sectors, and the $2$-fold degeneracy}, we discuss the implication of this anomaly in the magnetic sectors selected by  $\tilde{\mathbb Z}_{2(N\pm2)}^\chi$ and exhibit the exact degeneracy. In the Appendix, we show the equivalence of our operator construction to the ``half-gauging'' of a discrete subgroup of the magnetic one form symmetry $U(1)^{(1)}_m$, used in \cite{Choi:2022jqy,Cordova:2022ieu} to construct a properly normalized defect yielding consistent Euclidean correlation functions. We conclude with a brief discussion in Section \ref{Discussion}.  

%%%%%%%%%%%%%%%%%%%%%%%%%%%%%%%%%%%%%%%%%%%%%%%%%%%%%%%% 
\section{Noninvertible symmetries and their anomalies}
\label{gauge theories noninvertible symmetries}
 %%%%%%%%%%%%%%%%%%%%%%%%%%%%%%%%%%%%%%%%%%%%%%%%%%%%%%%%%
 
%Now, we consider gauging the $U(1)$ baryon number such that the full gauge group becomes $SU(N)\times U(1)$. 
Consider $SU(N) \times U(1)$ gauge theory with a single-flavor massless  Dirac fermion in a representation $R$. We use $n$, $T_R$, and $d_R$ to denote the $N$-ality, Dynkin index, and the dimension of $R$, respectively, and focus mostly on  $T_R = N \pm 2$ for the two-index symmetric (S)/antisymmetric (AS) representations of $N$-ality $n=2$ and dimension $d_R={N(N\pm1)\over 2}$ for S/AS. Yet, our construction can be easily generalized to theories with several flavors and fermions in higher representations. Classically, the theory is endowed with a $U(1)_\chi$ global chiral symmetry. The fermion charges (both are left-handed Weyl) under $(SU(N), U(1), U(1)_\chi)$ are 
 \begin{equation}
 \psi_R \sim (R, 1, 1), ~~ \psi_{\bar R} \sim (\bar R, -1, 1)\,.
 \end{equation}
  The theory with the gauged $U(1)$ has a $\mathbb Z_N^{(1)}$ $1$-form  electric symmetry, acting on both $SU(N)$ and $U(1)$ Wilson loops, as well as a $U(1)^{(1)}_m$  $1$-form magnetic  symmetry which distinguishes the different $U(1)$-flux sectors.
  
We use $A$ and $a$ for the $1$-form gauge fields of $SU(N)$ and $U(1)$, respectively. The corresponding field strengths are $F$ and $f$. The anomaly equation for the chiral $U(1)_\chi$ current is
 \begin{equation}\label{anomaly1}
 \partial_\mu j_\chi^\mu -  2 T_R \partial_\mu K^\mu_{SU(N)} - {2 d_R \over  8 \pi^2} \epsilon_{\mu\nu\lambda\sigma} \partial^\mu a^\nu \partial^\lambda a^\sigma =0,\end{equation} 
 where $K^\mu_{SU(N)}$ is the $SU(N)$ topological current. Its normalization is such that the integral of $K^0_{SU(N)} \equiv K^{CS}$ over a three-dimensional manifold changes by an integer under large gauge transformations. In other words, the operator \begin{equation}\label{operator1}
 e^{- i 2 \pi \int_{\mathbb T^3} d^3 x \; K^{CS}(A) }
 \end{equation}
 is invariant under large gauge transformations. This operator (which shifts the $\theta$-angle by $2\pi$) will be important in what follows.
 
When the theory is defined on a spatial manifold with nontrivial $2$-cycles, e.g., on $\mathbb T^3$, gauging $U(1)$ breaks  $\mathbb Z_{2 T_R}^{\chi}$ of the $SU(N)$ theory further down to $\mathbb Z^{\chi}_{\scriptsize 2\mbox{gcd}(d_R, T_R)}$. The easiest way to see that is by compactifying the time direction, so we consider the Euclidean version of the theory on $\mathbb T^4$. Under a chiral transformation, $(\psi_R, \psi_{\bar R})\rightarrow e^{i\alpha}(\psi_R, \psi_{\bar R})$, the measure changes by
\begin{eqnarray}\label{anomaly2}
\exp\left[i\alpha\left(2T_Rc_2(F)+2d_R c_2(f) \right)\right]\,,
\label{the general phase}
\end{eqnarray}
where  $c_2(F)\equiv \int_{\mathbb T^4} \frac{\tr F\wedge F }{8\pi^2}\in \mathbb Z$ and  $c_2(f)\equiv \int_{\mathbb T^4}\frac{f\wedge f}{8\pi^2}\in \mathbb Z$ are the second Chern classes of $SU(N)$ and $U(1)$. Then, demanding that the phase (\ref{the general phase}) is trivialized, and using B\'ezout's identity, which states that integers of the form $az_1+bz_2$ are exact multiples of $\mbox{gcd}(a,b)$, we arrive at our conclusion.\footnote{As a side note, there is no solution of self-dual BPST instanton on $\mathbb T^4$ with unit topological charge and zero twists \cite{Braam:1988qk}. Adding a twist removes the obstruction
to the existence of the solution. These twists are discussed below.} For example, when gcd($d_R, T_R)=1$, there is no residual chiral symmetry and the only symmetry left over is the $\mathbb Z_2$ fermion number symmetry which stays intact, assuming the Lorentz symmetry is unbroken.

%%%%%%%%%%%%%%%%%%%%%%%%%%%%%%%%%
%\subsection{The noninvertible chiral symmetry and its anomaly}
%%%%%%%%%%%%%%%%%%%%%%%%%%%%%%%%

It is important to emphasize that $\mathbb Z^{\chi}_{\scriptsize 2\mbox{gcd}(d_R, T_R)}$ is a genuine invertible symmetry, which is represented by a unitary operator acting on the Hilbert space of the theory. What is the fate of $\mathbb Z_{2T_{R}}^{\chi}$ that is broken because of the $U(1)$ instantons? Below, following the approach of \cite{Karasik:2022kkq,GarciaEtxebarria:2022jky}, we argue that this symmetry becomes noninvertible. We shall exhibit this and study the consequences using  Hamiltonian quantization on $\mathbb T^3$ in a very explicit manner.

Consider the Hamiltonian quantization of the theory on a rectangular $\mathbb T^3$ of sides $L_1, L_2, L_3$, in  the $A_0=a_0=0$ gauge.
The Hilbert space is constructed in terms of gauge fields $A = A_i dx^i$ and $a = a_i dx^i$, where $i=1,2,3$ is the spatial index (below, we use $\hat e_i$ to denote a unit vector in the $i$-th direction). We often use $x,y,z$ for $x_1,x_2,x_3$ when writing the components explicitly.  

The physical Hilbert space is obtained from field-operator eigenstates after appropriate gauge averaging and imposing Gauss's law; a detailed description can be found in \cite{Cox:2021vsa}, see also the earlier works  \cite{GonzalezArroyo:1987ycm, vanBaal:2000zc}. The gauge fields obey boundary conditions on $\mathbb T^3$
\begin{eqnarray}
 \nonumber 
A(x + \hat e_i L_i) &=& \Gamma_i A(x) \Gamma_i^{-1}\,,\\
a(x + \hat e_i L_i) &=& a(x) - d \omega_i(x)~, 
\label{gaugefields}
\end{eqnarray}
given in terms of $SU(N)$ and $U(1)$ transition functions, $\Gamma_i$ and $\omega_i$, respectively  \cite{tHooft:1979rtg}.  We work in a gauge where the $SU(N)$ transition functions $\Gamma_i$ are constant unitary $N\times N$ matrices \cite{Gonzalez-Arroyo:1997ugn}. 
The fermions obey similar boundary conditions (for brevity, we write these for the $N$-ality two case, $n=2$):
\begin{eqnarray}
\nonumber
\psi_R(x + \hat e_i L_i) &=&\Omega_i(x) \Gamma_i \; \psi_R(x) \; \Gamma_i^{T}\,,\\
\psi_{\bar R}(x + \hat e_i L_i) &=&\Omega_i(x)^{-1} \Gamma_i^* \; \psi_{\bar R}(x)\; \Gamma_i^{\dagger}.
\label{fermions}
\end{eqnarray} 
The $U(1)$ and $SU(N)$  transition functions obey cocycle conditions assuring that the fields  satisfying (\ref{gaugefields}, \ref{fermions}) are single-valued on the torus. 
The cocycle conditions obeyed by the transition functions are
\begin{eqnarray}
\nonumber
\Gamma_i \Gamma_j &=& e^{i {2 \pi \over N} n_{ij}}\; \Gamma_j \Gamma_i \nonumber ~, ~ n_{ij} = - n_{ji}, ~ n_{ij} \in \mathbb Z \; (\text{mod} N), \\
\Omega_i(x +\hat e_j L_j) \Omega_j(x) &=& e^{- i   {2 \pi n \over N} n_{ij}} \;\Omega_j(x + \hat e_i L_i) \Omega_i(x)\,,
\label{cocycles}
\end{eqnarray}
or
\begin{eqnarray}\label{omegau1}
\Omega_i(x) = e^{i \omega_i(x)},\; \text{with} ~ \omega_i =  \sum\limits_{j=1}^3 \pi(m_{ij} + {n \over N} n_{ij}) {x_j \over L_j}, ~ m_{ij} = - m_{ji} \in \mathbb Z\,.
\end{eqnarray}
Here, the (mod $N$) integers $n_{ij}$ represent topological classes of $2$-form $\mathbb Z_N^{(1)}$ background fields in the respective $2$-planes and $m_{ij}$ label integer $U(1)$-flux sectors, distinguished by their magnetic flux through the various 2-planes. The integer $m_{ij}$ are charges under the global magnetic $U(1)_m^{(1)}$ 1-form symmetry \cite{Gaiotto:2014kfa}. It is easily seen that all gauge and matter fields are single-valued on $\mathbb T^3$ when (\ref{cocycles}) are obeyed. The integers $n_{ij}$ and $m_{ij}$ label different flux sectors of the torus Hilbert space.

We next consider the global $\mathbb Z_N^{(1)}$ symmetry  \cite{Gaiotto:2014kfa}.
The generators of the $1$-form $\mathbb Z_N^{(1)}$ symmetry act on the transition functions. We label by $t_j$ and $T_j$ the $U(1)$ and $SU(N)$ group elements representing the action of the  generators of $\mathbb Z_N^{(1)}$ in the $j$-th direction on the $U(1)$ and $SU(N)$ transition functions, respectively. The action of   $\mathbb Z_N^{(1)}$ is given by
\begin{eqnarray}
\label{center}
\Omega_j &\rightarrow& t_i(x+\hat e_j L_j)\; \Omega_j(x) \;t_i^{-1}(x) =  e^{- i {2 \pi n \over N} \delta_{ij} } \Omega_j(x), \nonumber \\
\Gamma_j &\rightarrow&  T_i(x+\hat e_j L_j) \;\Gamma_j\; T_i^{-1}(x) =   e^{ i {2 \pi  \over N} \delta_{ij} } \Gamma_j . 
\end{eqnarray}
Put differently, $\mathbb Z_N^{(1)}$ transformations are represented by ``improper'' gauge transformations on the Hilbert space (in the first line, since both $t_j$ and $\Omega_i$ are abelian, the $\Omega_j$'s can be dropped).

The explicit expressions for $U(1)$ and $SU(N)$ group elements $t_j(x)$ and $T_j(x)$ obeying (\ref{center}) and generating the center symmetry can be worked out. The explicit form of $T_j(x)$ can be found in the literature\footnote{See \cite{GarciaPerez:1992fj,Poppitz:2022rxv} for $SU(2)$ and \cite{Selivanov:2000kg} for $SU(N)$.}. The explicit form of $T_j(x)$ depends on both the choice of gauge for the transition functions and on the $n_{ij}$ 2-form background (this is because $T_j (x)$ have nontrivial winding numbers for $n_{ij} \ne 0$ (mod$N$) \cite{tHooft:1981sps}).  We will not need the expression for $T_j$, but only the commutation relation of the $SU(N)$ $\mathbb Z_N^{(1)}$ center symmetry generators and the gauge invariant operator (\ref{operator1}),
\begin{equation}
\label{gaugeq3}
T_i \; e^{- i 2 \pi \int_{T^3} d^3 x K^{CS}(A)} \; T_i^{-1}  = e^{- i {2 \pi {\epsilon_{ijk} n_{jk} \over 2 N}} }\;  e^{- i 2 \pi i \int_{T^3} d^3  K^{CS}(A)}\,,
\end{equation}
which is nontrivial in the presence of a 't Hooft twist; see \cite{Cox:2021vsa} for a derivation. 
On the other hand, the center symmetry generators' action on the  $U(1)$   can be taken
\begin{eqnarray}\label{centeru1}
t_i(x) = e^{- i {2 \pi n \over N}{x_i \over L_i}} = e^{ i \tilde\lambda^{(i)}(x)}~, ~\text{with} ~ \tilde\lambda^{(i)}(x + \hat e_j L_j) = \tilde\lambda^{(i)}(x) - \delta^{ij} {2 \pi n \over N}~.
\end{eqnarray}

Next, we 
  construct the generator of the noninvertible chiral symmetry. To this end, we integrate the anomaly equation (\ref{anomaly1}) and use it to define a conserved (but not gauge-invariant) $U(1)_\chi$ symmetry operator on $\mathbb T^3$.  Because of the boundary conditions in the space directions for the $U(1)$ fields, whose transition functions (\ref{omegau1}) necessarily depend on $x_i$, there are $a$-dependent boundary terms.\footnote{On the other hand, since the  transition functions $\Gamma_i$ are constant, no such boundary terms appear for $SU(N)$.} We find,    denoting $K^0(a) \equiv  {1 \over 8 \pi^2}\epsilon^{ijk} a_i \partial_j a_k $:
\begin{eqnarray}\label{integratedanomaly}
0 &=& \int_{\mathbb T^3} d^3 x \left[ j_\chi^0 - 2 T_R K^{CS}(A) - 2 d_R K^0(a) \right]\bigg\vert^{x_0 = L_0}_{x_0=0} - {2 d_R \over 8 \pi^2} \int\limits_{0}^{L_0} dx_0 \int d^2 S^i  \epsilon^{ijk} a_j \partial_0 a_k\bigg\vert^{x_i = L_i}_{x_i=0}. \nonumber 
\\
\end{eqnarray}
Next, we note that $\partial_0 a_k$ is periodic on $\mathbb T^3$, while $a_k$ itself obeys (\ref{gaugefields}) with transition function $\omega_i$ of (\ref{omegau1}). It is easy to see that the second term above is also a total time derivative.  Integrating (\ref{integratedanomaly}),  we finally obtain that $Q_\chi(x_0=L_0)= Q_\chi(x_0=0)$, where
\begin{eqnarray}\label{q5}\nonumber
Q_\chi &=& \int_{\mathbb T^3} d^3 x \left[ j_\chi^0 - 2 T_R K^{CS}(A) - 2 d_R K^0(a) \right]\\
&&+  { d_R \over 4 \pi} (m_{xy} + {n \over N} n_{xy})\left[ \int\limits_{0}^{L_y} {d y \over L_y} \int\limits_{0}^{L_z} d z a_z(x=0,y,z) +\int\limits_{0}^{L_x} {d x \over L_x} \int\limits_{0}^{L_z} d z a_z(x,y=0,z)  \right]  \nonumber\\
&&+ \sum\limits_{\scriptsize\mbox{cyclic}}  (x \rightarrow y \rightarrow z \rightarrow x)\,.
\end{eqnarray}
The last line above indicates that there are two more terms obtained from the term on the second line by cyclic rotation of $x,y,z$.
Again, this is the operator $Q_\chi$ in the sector of Hilbert space with $U(1)$ fluxes $m_{ij}$ and $\mathbb Z_N^{(1)}$ fluxes $n_{ij} $. 

Exponentiating (\ref{q5}), we find 
the (non-gauge-invariant) operator representing $\mathbb Z_{2 T_R}^{\chi}$:
\begin{eqnarray}\label{x1}
X_{2 T_R} = e^{i {2 \pi \over 2 T_R} Q_\chi}~.
\end{eqnarray}
Let us now study the gauge transformation properties of   $X_{2 T_R}$. First, we note that because of the gauge invariance of (\ref{operator1}), the invariance of (\ref{x1}) under $SU(N)$ (large and small) gauge transformations is manifest. 
Next, consider  $U(1)$ gauge transformations with periodic\footnote{\label{footnote1}For use below, we also recall from (\ref{centeru1}) that the generator $t_j$ of the global $\mathbb Z_N^{(1)}$ in the $j$-th direction acts on the $U(1)$ gauge field as a nonperiodic gauge transformation, i.e. is obtained from (\ref{gaugeproper1}) upon replacing $n_j \rightarrow -{n\over N}$, where, we remind the reader, $n$ is the $N$-ality of the matter representation.}  $e^{i \lambda}$:
\begin{eqnarray}
\label{gaugeproper1}
a_i \rightarrow a_i -\partial_i  \lambda, ~\text{with} ~ \lambda(x + \hat e_i L_i) = \lambda(x) + 2 \pi n_i~.
\end{eqnarray}
To study the transformation properties of (\ref{x1}) under $U(1)$ transformations (\ref{gaugeproper1}), we note that,  with $\lambda$ from (\ref{gaugeproper1}),    $- 2\pi \int_{T^3}d^3 x K^0(a)$ transforms as
\begin{equation}\label{gaugeq1}
 - {1 \over 4 \pi} \int_{\mathbb T^3} d^3 x \epsilon^{ijk} a_i \partial_j a_k\bigg\vert^{a - d \lambda}_a = {n_x \over 2} \int dy dz (\partial_y a_z - \partial_z a_y) + \sum\limits_{\scriptsize\mbox{cyclic}}  (x \rightarrow y \rightarrow z \rightarrow x)~.
 \end{equation}
Since, recalling (\ref{gaugefields}, \ref{cocycles}),  
\begin{equation}\label{background1}
\int dy dz (\partial_y a_z - \partial_z a_y)= - 2 \pi (m_{yz} + {n n_{yz} \over N}),
\end{equation} we find
\begin{eqnarray}\nonumber\label{gaugeq11}
- 2 \pi \int_{\mathbb T^3} d^3 x K^0(a)\bigg\vert^{a - d \lambda}_a &=&  - {1 \over 4 \pi} \int_{T^3} d^3 x \epsilon^{ijk} a_i \partial_j a_k\bigg\vert^{a - d \lambda}_a = - \pi n_x  (m_{yz} + {n n_{yz} \over N})\\
 &&+ \sum\limits_{\scriptsize \mbox{cyclic}}  (x \rightarrow y \rightarrow z \rightarrow x)~.
 \end{eqnarray}
Likewise, we also find the $U(1)$ gauge transformation of the boundary terms in (\ref{q5})
\begin{equation}\label{gaugeq2}
 \int\limits_{0}^{L_y} {d y \over L_y} \int\limits_{0}^{L_z} d z a_z(x=0,y,z) +\int\limits_{0}^{L_x} {d x \over L_x} \int\limits_{0}^{L_z} d z a_z(x,y=0,z)\bigg\vert^{a - d \lambda}_a = - 4 \pi n_z~,
\end{equation}
along with two identical relations obtained by cyclic permutations of $x,y,z$.

Thus, combining the $U(1)$ gauge transformations (\ref{gaugeq11}, \ref{gaugeq2}), with the expression for $Q_\chi$ from (\ref{q5}), we find the transformation of $X_{2 T_R}$ 
 \begin{eqnarray}\label{xtransformu1}
 X_{2T_R}[a - d \lambda]  = X_{2 T_R}[a]\; e^{- i2 \pi  {d_R \over T_R}\left[  n_x (m_{yz}  + {n n_{yz} \over N}) +      n_y( m_{zx} +{n n_{zx} \over N})+ n_z (m_{xy}+{n n_{xy} \over N})\right]}\,,
 \end{eqnarray}
 showing explicitly that the operator (\ref{x1}) is not $U(1)$ gauge invariant.
 
 Thus, following \cite{Karasik:2022kkq,GarciaEtxebarria:2022jky}, to make $X_{2 T_R}$ gauge invariant,  we  sum over $n_x, n_y, n_z$, obtaining the noninvertible $\tilde{\mathbb Z}_{2T_R}^{\chi}$ operator 
 \begin{eqnarray}\label{tildex}
\tilde X_{2 T_R}  = e^{i {2 \pi \over 2 T_R} Q_\chi} \sum\limits_{n_x, n_y, n_z \in Z}e^{- i2 \pi  {d_R \over T_R}\left[  n_x (m_{yz}  + {n n_{yz} \over N}) +      n_y( m_{zx} +{n n_{zx} \over N})+ n_z (m_{xy}+{n n_{xy} \over N})\right]}\,,
\end{eqnarray}
with $Q_\chi$ from (\ref{q5}). This equation shows that the operator is noninvertible and determines the sectors not annihilated by $\tilde X_{2 T_R}$. To see this, we use the Poisson resummation formula
\begin{eqnarray}
\sum_{n_x \in Z} e^{- i 2 \pi {d_R \over T_R}(m_{yz} + {n n_{yz} \over N}) n_x} = \sum_{l_x \in Z} \delta\left( {d_R \over T_R}(m_{yz} + {n n_{yz} \over N}) - l_x\right)\,.
\end{eqnarray}
For $\tilde X_{2 T_R}$ to act nontrivially, i.e., not be set to zero, it must be that in each two-plane, the fluxes $m_{ij}, n_{ij}$, $i<j$, have to obey 
\begin{equation}\label{condition1}
{d_R \over T_R}(m_{yz} + {n n_{yz} \over N}) = l_x, ~ l_x \in \mathbb Z\;  \; (\text{plus cyclic})\,.
\end{equation}
It is easy to see that such integer-valued combinations of fluxes always exist (we shall see examples below). In the Appendix, we construct the operator $\tilde X_{2 T_R} $ using the ``half-gauging'' procedure of refs.~\cite{Choi:2022jqy,Cordova:2022ieu}.

To summarize, here we have constructed a symmetry operator $\tilde X_{2 T_R}$ of the noninvertible chiral symmetry. The operator of the noninvertible symmetry acts as a projection operator: it annihilates  sectors of the torus Hilbert space whose fluxes do not obey (\ref{condition1}) and acts as unitary operator in each flux sector obeying (\ref{condition1}).\footnote{In backgrounds with $m_{ij}, n_{ij}$ chosen to yield integer $l_i$ (\ref{condition1}), the operator can simply be defined by (\ref{x1}), since, as (\ref{xtransformu1}) shows, for such values of $l_i$, it is gauge invariant.}

We next study the commutator of the $\mathbb Z_N^{(1)}$ center symmetry transformation with the noninvertible $\tilde X_{2T_R}$. We denote the $\mathbb Z_N^{(1)}$ generator in the $x$-direction by  $T_x t_x$ (for brevity omitting both hats over operators and the tensor product sign, with $T_x$, $t_x$ obeying (\ref{center})). We  use the commutation relation of (\ref{gaugeq3}) of $T_i$ with the operator (\ref{operator1}), as well the $U(1)$ transformation law derived above, eqn.~(\ref{xtransformu1}),  with only $n_x$ nonzero and with  the integer $n_x$ replaced by $-{n \over N}$ (as we reminded the reader in footnote \ref{footnote1}). We find  
\begin{eqnarray} \label{anomaly}
T_x  t_x \; \tilde X_{2 T_R} \; ( T_x t_x)^{-1} &=&  \tilde X_{2 T_R} e^{- i 2 \pi \left[ {n_{yz} \over  N} - {d_R  \over T_R} {n \over N} (m_{yz} + {n {n_{yz} \over N})}\right] }  \\
&=&   \tilde X_{2 T_R} e^{- i 2 \pi \left[ {n_{yz} \over  N} -  {n \over N} l_x \right] } ~, ~ \text{with} \; l_x \; \text{from} \; (\ref{condition1}). \nonumber
\end{eqnarray}

Equation (\ref{anomaly}) and its two cyclically permuted versions, constitute our main result. It shows that---provided both the phase on the r.h.s. is nontrivial and $\tilde X_{2 T_R}$ is nonzero (i.e.~(\ref{condition1}) holds)---there is a mixed anomaly between the $\mathbb Z_N^{(1)}$ center symmetry and the noninvertible $\tilde{\mathbb Z}_{2T_R}^{\chi}$ chiral symmetry of the $SU(N)\times U(1)$ theory. As both operators act nontrivially in the chosen integer-$l_i$  $\mathbb T^3$ Hilbert space, the algebra (\ref{anomaly}) will be seen to imply exact degeneracies for any size $\mathbb T^3$.

In Section \ref{Hilbert space, magnetic sectors, and  the $2$-fold degeneracy}, we shall show that there are cases where both the phase in (\ref{anomaly}) and $\tilde X_{2 T_R}$ are nontrivial, i.e., (\ref{condition1}) holds in the appropriate Hilbert space. As already familiar from   \cite{Cox:2021vsa},  this will be seen to imply degeneracies in the Hilbert space between different ``electric flux'' states (eigenstates of $\mathbb Z_N^{(1)}$, i.e., of  $T_x t_x$).

Before we continue with analyzing the implications of (\ref{anomaly}) for the finite-volume spectrum, let us make a connection of (\ref{anomaly}) with the Euclidean path integral. We denote  the $\mathbb T^3$ Hilbert space by ${\cal H}_{l_i}$,  with the understanding that  $U(1)$ ($m_{ij}$) and $\mathbb Z_N^{(1)}$ ($n_{ij}$) fluxes are chosen to yield an integer-$l_i$, so that $\tilde X_{2T_R}$ acts nontrivially. We now  define the $T_x t_x$-twisted partition function via the Hamiltonian formalism as a trace\footnote{The relations we derive below hold also if we insert $(-1)^F$ in the partition function.} over states in ${\cal H}_{l_i}$ with a $\mathbb Z_N^{(1)}$  $x$-direction generator inserted in the partition function, i.e.~${\cal Z}\equiv \tr_{{\cal H}_{l_i}} e^{- \beta H} T_x t_x$. Then, we use the fact that $\tilde{X}_{2 T_R}$ acts as an invertible unitary operator in  ${\cal H}_{l_i}$, as well as the commutation relation (\ref{anomaly}), to obtain
\begin{eqnarray}
\nonumber
\label{twistedZ}
{ \cal Z}&=& \tr_{{\cal H}_{l_i}}\left[ e^{- \beta H} T_x t_x\right] =  \tr_{{\cal H}_{l_i}} \left[e^{- \beta H} T_x t_x \tilde{X}_{2 T_R} \tilde{X}^\dagger_{2 T_R}\right]\\
 &=& e^{- i 2 \pi \left[ {n_{yz} \over  N} -  {n \over N} l_x \right] }\; \tr_{{\cal H}_{l_i}}\left[ e^{- \beta H} T_x t_x\right]\,.
\end{eqnarray}
We conclude that the chiral symmetry $\tilde X_{2 T_R}$ implies that ${\cal Z} = e^{- i 2 \pi \left[ {n_{yz} \over  N} -  {n \over N} l_x \right] } {\cal Z} $, so that, if the phase is nontrivial, ${\cal Z}$ vanishes, unless fermion fields $(\psi_{\bar R} \psi_R)^k$ are inserted to make ${\cal Z} $ nonzero.\footnote{Below, we show that  the phase  in  (\ref{twistedZ})  can take values at most in $\mathbb Z_2$, and that $k = {N \pm 2 \over 2}$, for S/AS fermions, for the values of $N$ and choice of fluxes where the phase is nontrivial.} To obtain a path integral interpretation of (\ref{twistedZ}), we note that the twisted partition function (\ref{twistedZ}) sums over $SU(N)$ and $U(1)$ gauge fields which obey the boundary conditions (\ref{gaugefields}) in the $\mathbb T^3$ spatial directions, determined by $n_{ij}$ and $m_{ij}$. The boundary conditions in the Euclidean time direction (of extent $\beta$) are twisted, by the insertion of $T_x$, leading to a nontrivial $SU(N)$ twist $n_{x4}=1$ in the $x$-time plane. Thus, the twisted partition function (\ref{twistedZ}) sums over $SU(N)$ field configurations with topological charges $ -{n_{14} n_{yz} \over N} +k = -{n_{yz} \over N} + k$, with all possible integer $k$ \cite{vanBaal:1982ag}. On the other hand, the insertion of $t_x$ implies that the $U(1)$ background obeys $a(x + \hat e_{4} \beta) = a (x) +  {2 \pi n \over N}{d x \over L_1}$ and thus $f_{14} = -{2 \pi n \over N L_1 L_4}$.  Recalling that the $U(1)$ field strength in the $yz$ plane is ${2 \pi \over L_2 L_3} (m_{yz} + {n \over N} n_{yz})$), the $U(1)$ topological charge  is then  equal to $ { n\over N}(m_{yz} + {n \over N} n_{yz})$. Applying a chiral transformation with $\alpha = {2 \pi \over 2 T_R}$,  and using the measure transformation (\ref{anomaly2}) (with $c_2(F)$, $c_2(f)$ substituted by the fractional topological charges just mentioned) we obtain the phase  $e^{- i 2 \pi [{n_{yz} \over N} - {d_R \over T_R}{n \over N} (m_{yz} + {n \over N} n_{yz})]}$, which, after using (\ref{condition1}), is seen to be the same as in (\ref{twistedZ}), as expected.
 
%%%%%%%%%%%%%%%%%%%%%%%%%%%%%%%%%%%%%%
\section{Hilbert space, magnetic sectors, and  the $2$-fold degeneracy}
\label{Hilbert space, magnetic sectors, and  the $2$-fold degeneracy}
%%%%%%%%%%%%%%%%%%%%%%%%%%%%%%%%%%%%%%%

Here, we analyze the consequences of condition (\ref{condition1}) and the algebra of (\ref{anomaly}). Condition (\ref{condition1})  selects sectors in Hilbert space with definite $U(1)$ magnetic fluxes in the $2$-$3$, $3$-$1$, and $1$-$2$ planes proportional to the integers $l_x, l_y, l_z$, respectively. In the following, we always set $l_y=l_z=0$ (with $n_{xy}=m_{xy}=n_{zx}=m_{zx}=0$)  to reduce complexity and examine the theory for various values of $l_x={d_R \over T_R}(m_{yz} + {n n_{yz} \over N})$. We also choose $n_{yz}\in \{0,1,2,..., N-1\}$. Since we are mainly concerned with symmetric (S)/antisymmetric (AS) fermions, we set the $N$-ality $n=2$. 

We start with sectors with a vanishing magnetic flux, i.e., we set $l_x=0$, which translates into $m_{yz}+\frac{2}{N}n_{yz}=0$.  When $N$ is odd, the only solution is the null solution $m_{yz}=n_{yz}=0$. This yields a trivial phase in the algebra of  (\ref{anomaly}). However, when $N=2M$ is even, there are two solutions. First, the null one $m_{yz}=n_{yz}=0$, giving a trivial phase in (\ref{anomaly}). The second is the new solution (we denote the phase in (\ref{anomaly}) by $e^{- i \alpha} = e^{- 2 \pi i  ({n_{yz} \over N} - {2\over N} l_x)}$):
\begin{eqnarray}\label{zeroflux}
\mbox{S/AS}: && N=2 M, ~ n_{yz}={M}, ~ m_{yz}=-1,~ l_x = 0,  \alpha =  \pi.  
\end{eqnarray}
The   $\mathbb Z_2$ phase in (\ref{anomaly})  implies that some electric flux states in a sector with a $0$-magnetic flux are $2$-fold degenerate.

Numerical tests reveal that this pattern continues in sectors with nonzero magnetic flux, $l_x\neq 0$. First, when $N$ is odd, all allowed sectors have a trivial phase in the algebra of (\ref{anomaly}), indicating no kinematical constraints in these theories.  For $N$ even, $N=2M$, one can use $l_x={d_R \over T_R}(m_{yz} + {n n_{yz} \over N})$  to see that there exists integers $m_{yz} \in \mathbb Z$ and $n_{yz}\in \{0,1,2,..., N-1\}$  that satisfy the relation
\begin{eqnarray}
M\left(2M\pm1)(Mm_{yz}+n_{yz}\right)=2M(M\pm1)l_x\,,
\label{l sectors condition 1}
\end{eqnarray}
for S/AS fermions, respectively. Sectors with definite $m_{yz}$ and $n_{yz}$ that satisfy (\ref{l sectors condition 1}) will also yield at most a $\mathbb Z_2$ phase in the algebra (\ref{anomaly}), if they additionally satisfy
\begin{eqnarray}
M^2\left(n_{yz}+(2M\pm1)m_{yz} \right)\in M^2(M\pm 1)(2\mathbb Z+1)\,.
\label{l sectors condition 2}
\end{eqnarray}
The $\mathbb Z_2$ phase implies that the states in these sectors have double-fold degeneracy. 

Let us flesh this out in detail. For definiteness,  we consider S/AS fermions, which yield a nontrivial phase in (\ref{anomaly}) as well as involve only a minimal $SU(N)$ 't Hooft twist, $n_{yz}=1$ (as opposed to the $l_x=0$ solutions (\ref{zeroflux}), which must have $n_{yz}=N/2$ to produce an anomaly). For both S/AS fermions, these minimal-twist solutions, yielding an integer $l_x$ and a $\mathbb Z_2$ phase in (\ref{anomaly}), must have $N=4p+2$. Using again $e^{- i \alpha} = e^{- 2 \pi i  ({n_{yz} \over N} - {2\over N} l_x)}$, examples of $m_{yz}$ that give the $
\mathbb Z_2$ phase are:
\begin{eqnarray}\label{nonzeroflux}\nonumber
\mbox{S}: && N=4p+2, ~ n_{yz}=1, ~ m_{yz}=-2p-1,~ l_x = -p(3 +4 p),  \alpha = 2\pi( {1 \over 2} + 2 p),  \\
\nonumber
\mbox{AS}: && N=4p+2, ~ n_{yz}=1,~ m_{yz}=-2p-1, ~ l_x = -(1+p)(1+4p),\alpha = 2 \pi({3 \over 2} + 2 p)\,.\\
\end{eqnarray}

Thus, we now focus on the flux backgrounds (\ref{nonzeroflux}), enumerate the degenerate ``electric flux'' sectors\footnote{These states can be explicitly constructed in the semiclassical limit of a small $\mathbb T^3$ \cite{Witten:1982df,vanBaal:2000zc} (or a small $\mathbb T^2 \in \mathbb T^3$, with the $\mathbb T^2$ spanning the $y,z$ directions \cite{Tanizaki:2022ngt,Tanizaki:2022plm})  by focusing on the lowest energy states. We shall not do this here in complete detail, but see Footnote \ref{abc}.} in the corresponding Hilbert spaces ${\cal{H}}_{l_x, l_y=l_z = 0}$, and discuss some of their properties. The transition functions for $SU(N)$ and $U(1)$ obeying the cocycle conditions with $n_{yz}, m_{yz}$ given in (\ref{nonzeroflux}) are
 \begin{eqnarray}
 \Gamma_x &=& 1,~ \Gamma_y = P,~\Gamma_z = Q,~\text{where}~ P Q = e^{i {2\pi \over N}} Q P~, \nonumber \\
 \omega_x &=&1, ~\omega_y = -{\pi z \over L_3}{4p(p+1)\over 2p+1},~ \omega_z = {\pi y \over L_2}{4p(p+1)\over 2p+1}~.
 \end{eqnarray}
The center symmetry generators for $SU(N)$ and $U(1)$, obeying (\ref{center}), can be taken to be\footnote{For $SU(N)$, these are the ones from \cite{Witten:1982df}. Briefly, we remind the reader that $T_x$ cannot be taken to be constant, since, as already discussed, a twist of the partition function in the time direction by $T_x$, recall (\ref{twistedZ}),  leads to fractional topological charge on the $\mathbb T^4$, equal to ${1 \over N} + {\rm{integer}}$. This implies that $T_x$ has fractional winding number \cite{tHooft:1981sps}, as a map from $\mathbb T^3$ to the gauge group, with its $N$-th power being a large gauge transformation. Thus, on physical states $T_x$ obeys $T_x^N = 1$. Explicit expressions for $T_1$ can be found in the literature (see \cite{GarciaPerez:1992fj,Poppitz:2022rxv} for $SU(2)$ and \cite{Selivanov:2000kg} for $SU(N)$) but are not needed here.}
  \begin{eqnarray}
  T_x &=& T_x(y,z) ~ (\mbox{recall}\,,T_x^N |{\rm{phys}}\rangle = |{\rm{phys}} \rangle), T_y = Q^{-1}, T_z = P,  \nonumber \\
   t_k &=& e^{- i {2 \pi \over 2p+1} {x_k \over L_k}}~. 
  \end{eqnarray}
  Since $T_x t_x$ is a symmetry, eigenstates of the Hamiltonian can be labeled by its eigenvalues, $T_x t_x |E, e_x\rangle = |E, e_x\rangle e^{i {2 \pi \over N} e_x}$, with $e_x \in \mathbb Z$ (mod$N$). On the other hand, the algebra (\ref{anomaly}), $T_x t_x \tilde X_{2 T_R} (T_x t_x)^{-1} = - \tilde X_{2 T_R}$,   implies that, in ${\cal{H}}_{l_x, l_y=l_z=0}$, with $l_x$ from (\ref{nonzeroflux}),
 \begin{eqnarray}\label{degeneracy}
\tilde X_{2 T_R} |E, e_x\rangle_{{\cal{H}}_{l_x, l_y=l_z =0}} \sim |E, e_x + {N \over 2} ({\rm mod}\, N) \rangle_{{\cal{H}}_{l_x, l_y=l_z =0}}.
\end{eqnarray}
Thus, $\tilde X_{2 T_R}$ maps an eigenstate of the Hamiltonian of energy $E$ and flux $e_x$ to another eigenstate of the same energy, but with flux $e_x + {N\over 2}$ (mod$N$) (hence, a $T_x t_x$ eigenvalue differing by $e^{i\pi}$, as per (\ref{anomaly}); we also note that the phase in the action of $\tilde X$ also depends on whether the state is bosonic of fermionic).\footnote{\label{abc}As promised, on a small $\mathbb T^3$, the lowest flux states can be worked out classically. For the bosonic backgrounds, the lowest-energy gauge field backgrounds are 
\begin{eqnarray}\label{background}
A^{(l)} &=& - i T_1^l d T_1^{-l}, ~l=0,...,N-1, \nonumber  \\
a^{(l)} &=& {4 \pi p(p+1) \over L_3 L_2 (2p+1)}(y dz - z dy) + {2 \pi l \over (2 p + 1) L_1}  dx\,.
\end{eqnarray}
The fundamental $SU(N)$ winding Wilson loops then take values $W_{y} = W_z = 0, W_x = e^{i {2 \pi \over N} l}$. Using the backgrounds (\ref{background}), solving for the fermions, imposing Gauss's law,  and averaging over gauge transformations, one can construct the $N$  classically-degenerate states in ${\cal{H}}_{l_x, l_y=l_z=0}$. It is already clear from (\ref{background}) that these $N$ states are obtained by the action of $T_x t_x$ from each other. The electric flux states from eqn.~(\ref{degeneracy}) are a discrete Fourier transform thereof. The anomaly implies that the $N$-fold degeneracy will be lifted and that only the pairwise degeneracy will remain quantum mechanically. Extending this small-torus explicit analysis further, along the lines of \cite{Tanizaki:2022ngt,Tanizaki:2022plm}, as well as similar studies for other backgrounds, e.g., ~(\ref{zeroflux}), are left for the future.
}

To further characterize the degenerate flux states, 
we will show that the degenerate states (\ref{degeneracy}) have nonvanishing expectation values of a condensate, which we write schematically as $(\psi_{\bar R} \psi_R)^{N \pm 2\over 2}$. These expectation values take opposite values in the two degenerate flux states.
To this end, we now go back to our twisted partition function (\ref{twistedZ}). Since in ${\cal{H}}_{l_x,l_y=l_z=0}$ the phase is in $\mathbb Z_2$, in order to obtain a nonzero phase we must  insert $(\psi_{\bar R} \psi_R)^{N \pm 2\over 2}$,  a gauge invariant object which transforms with a $\mathbb Z_2$ phase under $\psi_{R, \bar R} \rightarrow e^{i {2 \pi \over 2 (N\pm 2)}} \psi_{R, \bar R}$. We have 
\begin{eqnarray}\label{condensate1}
\nonumber
\langle (\psi_{\bar R} \psi_R)^{N \pm 2\over 2}\rangle &=& \tr_{{\cal H}_{l_i}}\left[ e^{- \beta H}  (\psi_{\bar R} \psi_R)^{N \pm 2\over 2} T_x t_x\right]  \\
&=& \sum_{E,e_1=0,...,N-1} e^{- \beta E(e_1)} e^{i {2 \pi \over N}e_1} \langle E, e_1| (\psi_{\bar R} \psi_R)^{N \pm 2\over 2} |E, e_1\rangle\,.
\end{eqnarray}
Next, recall that $\tilde X_{2T_R}^\dagger \tilde X_{2 T_R} = 1$  in ${\cal{H}}_{l_i}$, remembering that we are in a definite magnetic flux sector, where the chiral symmetry operator $\tilde X_{2T_R}$ acts in an invertible manner, and using (\ref{degeneracy}), we find that   gauge-invariant  condensates obey
\begin{eqnarray}\label{condensate2}
\langle E, e_1| (\psi_{\bar R} \psi_R)^{k} |E, e_1\rangle = e^{i {2 \pi \over N \pm 2} k} \; \langle E, e_1 +{N \over 2} | (\psi_{\bar R} \psi_R)^{k} |E, e_1 + {N\over 2}\rangle~.
\end{eqnarray}

In particular, (\ref{condensate2}) shows that  the condensate appearing in (\ref{condensate1})
takes opposite values in the degenerate states. In the twisted partition function (\ref{condensate1}), this minus sign is cancelled by change of the phase $e^{i {2 \pi \over N}e_1}$ (from the action of $T_x t_x$). Thus, we can restrict the evaluation of (\ref{condensate1}) by summing over half the $e_1$ sectors:
\begin{eqnarray}\label{condensate3}
\langle (\psi_{\bar R} \psi_R)^{N \pm 2\over 2}\rangle &=& 2 \sum_{E,e_1=0,...{N\over 2}-1}   e^{- \beta E(e_1)}  e^{i {2 \pi \over N}e_1} \langle E, e_1| (\psi_{\bar R} \psi_R)^{N \pm 2\over 2} |E, e_1\rangle  ~.
\end{eqnarray}
These nonzero expectation values can be computed semiclassically  at a small torus and shown not to vanish, similar to \cite{Anber:2022qsz}.

To summarize, above we constructed the doubly-degenerate states using the background (\ref{nonzeroflux}), in the $N=4p+2$ theory, as an example. However, based on the $\mathbb Z_2$-valued anomaly,  similar descriptions involving degenerate states with opposite values of the relevant condensate hold in all even-$N$ cases. In particular, the doubly-degenerate flux states corresponding to (\ref{zeroflux}) can also be explicitly worked out.

Before we discuss these cases, let us contrast the findings in the $SU(N) \times U(1)$ theory on $\mathbb T^3$ with those in the $SU(N)$ theory, also on $\mathbb T^3$. Consider the $SU(4 p+2)$ theory (i.e., with even $N$ not divisible by $4$). It has a $\mathbb Z_2^{(1)}$ center symmetry and an invertible discrete chiral symmetry $\mathbb Z_{2(N\pm 2)}^\chi$. These have a $\mathbb Z_2$-valued mixed anomaly in appropriate $\mathbb Z_2^{(1)}$ backgrounds on $\mathbb T^3$. This anomaly implies, as in \cite{Cox:2021vsa}, an exact two-fold degeneracy in the twisted Hilbert space of the $SU(4p+2)$ theory, on any torus size. This is similar to the degeneracy of the $SU(4p+2) \times U(1)$ theory on $\mathbb T^3$ discussed in this paper. We stress, however, that the latter theory has gcd$(d_R, T_R) =1$, and hence no genuine chiral symmetry. The degeneracy we found is, thus, due to the noninvertible chiral $\tilde {\mathbb{Z}}_{2 T_R}^\chi$ symmetry.

Consider now the case when $N$  is divisible by $4$.  In the $SU(N)$ theory this mixed anomaly is trivial on $\mathbb T^3$ and hence one cannot use the $\mathbb Z_2^{(1)}$ $1$-form symmetry to argue for an exact degeneracy on a finite-size torus. In the $SU(N) \times U(1)$ theory, however, for any even $N$, we showed that  there is an  anomaly between the $\mathbb Z_N^{(1)}$ center and noninvertible chiral symmetries in the (\ref{zeroflux}) background, leading to an exact two-fold degeneracy at any size $\mathbb T^3$.

The case of minimal dimension condensate occurs if we take $N=4$ and an antisymmetric tensor. Here,  the anomaly predicts equal and opposite values of the bilinear fermion condensate $\psi_{\bar R} \psi_R$ in the two states that are degenerate at any finite volume, in the appropriately twisted background.
Since the degeneracy is present at any finite volume, should the condensate remain nonzero in the infinite volume limit, this predicts the $\mathbb Z_4 \rightarrow \mathbb Z_2$ (noninvertible) chiral symmetry breaking in the thermodynamic limit. 

 We stress that the use of appropriate twists---the $m_{ij}, n_{ij}$ with integer $l_i$, i.e. the ones that reveal the anomaly---at finite volume  is simply a tool to probe the gauge dynamics. At least in the theory with a nonzero mass gap, the infinite volume limit is expected to be independent of the boundary conditions and the degeneracies revealed are expected to persist in the thermodynamic limit.

%%%%%%%%%%%%%%%%%%%%%%%%%%%%%
\section{Discussion}
\label{Discussion}
 %%%%%%%%%%%%%%%%%%%%%%%%%%%%
 
Here, we found that the anomaly establishes the $2$-fold  degeneracy on arbitrary-size $\mathbb T^3$. Yet, one eventually wants to see what happens as we take the thermodynamic limit by sending the volume of $\mathbb T^3$ to infinity.

 To speculate on what could happen in $SU(N)\times U(1)$ theory, let us again return to its cousin, the $SU(N)$ gauge theory with S/AS fermions. The latter has a global $U(1)$ baryon number and $U(1)_\chi$ chiral symmetry. As usual, quantum effects  break $U(1)_\chi$ down to, now, the invertible $\mathbb Z_{2(N\pm2)}^\chi$. The $0$-form faithful global symmetry of this theory is $\frac{U(1)}{\mathbb Z_{N\over p}}\times \mathbb Z_{2(N\pm2)}^\chi$, with $p=\mbox{gcd}(N,2)$. The fact that the quotient group is nontrivial means that we can activate a 't Hooft flux in the center of $SU(N)$ accompanied by a flux in $U(1)$ such that the cocycle conditions are always obeyed on general four-dimensional manifolds.  The authors constructed these fluxes in vector-like theories, dubbed as the baryon-color (BC) fluxes, in \cite{Anber:2019nze,Anber:2020gig} (also see \cite{Anber:2021lzb,Anber:2020xfk,Anber:2020qzb,Lohitsiri:2022jyz,Nakajima:2022jxg} for applications, and \cite{Anber:2021iip}  for the construction and applications of these fluxes in chiral gauge theories).  The partition function acquires a $\mathbb Z_{N\pm 2}$ phase as we apply a $\mathbb Z_{2(N\pm2)}^\chi$ rotation in the background of the BC flux. This phase was interpreted as an anomaly of $\mathbb Z_{2(N\pm2)}^{\chi} $ in the BC background. Assuming that the theory has a mass gap and forms hadrons in the IR, the anomaly is   interpreted to imply the existence of $N\pm2$ degenerate vacua  \cite{Anber:2021lzb}.  One expects to see this degeneracy emerge on a finite-volume manifold (larger than the inverse strong-scale) and persist in the thermodynamic limit. 

As we argued in this work, gauging the $U(1)$ baryon symmetry endows the theory with a $\mathbb Z_2$ anomaly phase when $N$ is even, implying that at any finite volume there is exact $2$-fold degeneracy. Thus, $2$ degenerate vacua are guaranteed to survive the infinite volume limit. If the $SU(N) \times U(1)$ theory also has  $N \pm 2$ degenerate vacua in the infinite volume limit (as the $SU(N)$ theory is believed to) the exact $N \pm2$-fold degeneracy should be revealed in the thermodynamic limit. 

One might, of course, wonder whether a stronger phase (and stronger constraints, as in  \cite{Cordova:2019jqi}) can be exhibited if we subject the $SU(N)\times U(1)$ theory to a gravitational background. In other words, it would be interesting to investigate whether there is a mixed anomaly between the noninvertible   $\tilde{\mathbb Z}_{2(N\pm2)}^\chi$ and gravity.
If the gravitational anomaly does not produce a stronger phase beyond $\mathbb Z_2$, the implications in the thermodynamic limit of the $SU(N)\times U(1)$ theory are of interest. By weakly gauging $U(1)$ in the $SU(N)$ theory, one anticipates the presence of $N\pm2$ nearly degenerate vacua. However, the dynamics of the $U(1)$ gauge field may affect the precise degeneracy. This intriguing investigation remains open for future study.

Finally, we comment on the $SU(N) \times U(1)$ theory at finite temperature. The presence of the identified mixed anomaly implies that it is necessary for either the $1$-form symmetry $\mathbb Z_N^{(1)}$, the $0$-form symmetry $\tilde{\mathbb Z}_{2(N\pm2)}^\chi$, or both symmetries to be broken \cite{Gaiotto:2017yup,Shimizu:2017asf,Komargodski:2017smk}. Actually, since $\mathbb Z_N^{(1)}$ acts on the $U(1)$ Wilson lines, we expect it to be broken at zero and finite temperature. Consequently, the anomaly is consistently matched, leading us to only expect the restoration of the broken $\tilde{\mathbb Z}_{2(N\pm2)}^\chi$ symmetry and the breaking of the $\mathbb Z_2^{(1)}$ subgroup of the 1-form symmetry for even-$N$ at some finite temperature.

%%%%%%%%%%%%%%%%%%%%%%%
{\bf {\flushleft{Acknowledgments:}}} M.A. is supported by STFC through grant ST/T000708/1.   E.P. is supported by a Discovery Grant from NSERC. We thank Aleksey Cherman, Theo Jacobson, and Emily Nardoni for discussions. We also thank an anonymous referee for the suggestion to consider the ``half-gauging'' construction of the noninvertible defect and study its relation to the Hilbert space construction, see Appendix \ref{appx:defect}.
%%%%%%%%%%%%%%%%%%%%%%%

\appendix

\section{The noninvertible defect via  ``half-gauging'' of   $\mathbb{Z}_{T_R}^{(1)} \subset U(1)^{(1)}_m$}
\label{appx:defect}

Here, we consider the construction of the noninvertible defect for\footnote{Once again, we acknowledge our abuse of notation: as per the remark of Footnote \ref{fermionnumberfootnote}, the $\mathbb Z_2$ fermion number subgroup of ${\mathbb Z}_{2T_R}^\chi$ is part of the $U(1)$ gauge group.} $\tilde{\mathbb Z}_{2T_R}^\chi$ using the ``half-gauging'' procedure of refs.~\cite{Choi:2022jqy,Cordova:2022ieu} instead of the sum over gauge orbits of \cite{Karasik:2022kkq,GarciaEtxebarria:2022jky}. The advantage of the former construction is that it gives rise to well-defined Euclidean correlation functions involving the noninvertible defect, considered more generally than as an operator  inserted at a particular time. In particular, this allows  inserting the defect at a particular location in space, giving rise to well-defined Hilbert spaces twisted by the noninvertible symmetry.

Our discussion below makes use of the  techniques described explicitly in ref.~\cite{Choi:2022jqy} and is  restricted to the case gcd$(d_R,T_R) =1$. For our $SU(N) \times U(1)$ theories with a two-index S/AS Dirac fermion, this is the case of even-$N$ not divisible by $4$, leaving the generalization to the more general case for future work (although a generalization to gcd$(d_R,T_R)>1$ should be possible \cite{Choi:2022jqy}). In order to define the defect, we consider the gauging of the $\Z_{T_R}^{(1)}$ subgroup of the magnetic $U(1)_m^{(1)}$ 1-form symmetry. As shown in \cite{Choi:2022jqy},   this gauging  produces the same theory but with a discrete shift of the $U(1)$ theta angle $\theta \rightarrow \theta - {2 \pi d_R \over T_R}$. This shift, as per our Eqn.~(\ref{anomaly2}), is    undone by a  $\mathbb Z_{2 T_R}$ chiral rotation of the fermions $(\psi_R, \psi_{\bar R})$. We conclude that the $SU(N) \times U(1)$ theory is invariant under the above gauging of $\mathbb{Z}_{T_R}^{(1)} \subset U(1)_m^{(1)}$. 

As in \cite{Choi:2022jqy,Cordova:2022ieu}, the upshot of the half-gauging procedure is to define a defect,  replacing our eqn.~(\ref{tildex}) for the operator $\tilde X_{2 T_R}$ 
 by the following object, which we label, for brevity, by the same letter
\begin{eqnarray}
\label{qxx} 
\tilde X_{2 T_R}   &=& e^{ i{2 \pi \over 2 T_R} \int\limits_{t=0} d^3 x   j_\chi^0 -  i 2\pi \int\limits_{t=0} d^3 x  K^{CS}(A)  +    \int\limits_{t=0} {\cal A}^{T_R, d_R}[{d a \over T_R}] }~,
\end{eqnarray}
where, as in the main text,  $a$ is the dynamical $U(1)$ gauge field and $A$ is the $SU(N)$ gauge field with $K^{CS}[A]$ entering as in (\ref{operator1}). 

The 3d defect TQFT ${\cal A}^{T_R, d_R}[{d a \over T_R}]$  is defined via an integral over a 4d bulk with a boundary, which is here taken to be the $t=0$ plane:
\begin{eqnarray}\label{atheory}
 e^{ \;\; \int\limits_{t=0}  {\cal A}^{T_R, d_R}[{d a \over T_R}] }  &=& \int {\cal D}(b,c) \; e^{\;\; \int\limits_{t \ge 0}  \left({i d_R \over 4 \pi T_R}\; da \wedge da + {i \over 2 \pi }\; b^{(2)} \wedge da + { i T_R \over 2 \pi } \;b^{(2)}\wedge d c^{(1)} + {i k T_R  \over 4 \pi} \;b^{(2)}\wedge b^{(2)}\right) }. \nonumber \\ 
 &&
\end{eqnarray} 
The fields  $(b^{(2)}, c^{(1)})$   define the $2$-form $\mathbb Z_{T_R}$ gauge field (used in the half-gauging procedure) and obeying the Dirichlet boundary condition  $b^{(2)} =0$ at $t=0$, and $k$ is the modular inverse of $d_R$, i.e. $k d_R = 1 \;({\rm mod} T_R)$.  
Defining the defect via the ``half-gauging'' procedure  replaces the sum over gauge copies of   the gauge noninvariant $U(1)$ Chern-Simons term by the TQFT  ${\cal A}^{T_R, d_R}[{da\over T_R}]$ and produces a well defined Euclidean defect. 

The ${\cal A}^{T_R, d_R}[{da\over T_R}]$ theory has a $\Z_{T_R}^{(1)}$ global symmetry with an anomaly $d_R$, and the partition function $e^{ \;\int\limits_{t=0}{\cal A}^{T_R, d_R}[{da\over T_R}]}$ of  (\ref{atheory}) has the transformation properties of $e^{-i {d_R  T_R \over 4 \pi} \int\limits_{t \ge 0} {d a \over T_R} \wedge {da\over T_R}}$. In particular, under regular gauge transformations ${\delta a = d \omega}$ with $\oint d \omega = 2 \pi \Z$, 
\begin{eqnarray}
\label{deltaA}
\delta \;  e^{ \;\int\limits_{t=0}{\cal A}^{T_R, d_R}[{da\over T_R}]}=e^{i {d_R    \over   T_R} \int\limits_{t=0 } {d \omega} \wedge {da \over 2 \pi}}\;  e^{ \;\int\limits_{t=0}{\cal A}^{T_R, d_R}[{da\over T_R}]}~,
\end{eqnarray}
which implies, together with the gauge invariance of ${\cal A}^{T_R, d_R}[{da\over T_R}]$, that the defect vanishes unless ${d_R    \over   T_R} \oint da = 2\pi \Z$. This is precisely the condition (\ref{condition1}) of the main text,  in the absence of an electric $\Z_N^{(1)}$ center symmetry background.  Thus, we have succeeded in defining a noninvertible defect associated with the $\tilde{\mathbb Z}_{2T_R}^\chi$ chiral symmetry. 

The question left open, then, is to show that the definition of $\tilde X_{2 T_R}$ from (\ref{qxx}) reproduces the mixed anomaly with center symmetry exhibited in Eqn.~(\ref{anomaly}). This would  necessitate generalizing the half-gauging procedure to backgrounds with fractional  $\oint da \in {2 \pi \Z \over N}$.\footnote{Thus, incorporating the electric $\Z_N^{(1)}$ center-symmetry  background, denoted by $n_{yz}$ in (\ref{condition1}, \ref{anomaly}).} In the language of defects, one has to consider  defects associated with the (invertible) electric $\Z_N^{(1)}$ 1-form symmetry of the $SU(N) \times U(1)$ theory and determine the fusion rules of  these  codimension-two defects with the noninvertible defect defined by (\ref{qxx}). The fusion rules of $\tilde X_{2 T_R}$ with these codimension-two $\Z_N^{(1)}$-defects  of two different orientations are relevant to the anomaly (\ref{anomaly}). The first are  ``perpendicular''  to $\tilde X_{2 T_R}$, with a worldvolume in the $x$-$t$ plane and   in the language of this paper correspond  to the turning on  of $n_{yz}$  from (\ref{anomaly}). The other type of codimension-two  defects implementing the $\Z_N^{(1)}$ symmetry are ``parallel'' to $\tilde X_{2 T_R}$, with world volume in the $y$-$z$ plane,\footnote{We note that examples of fusion rules for similar parallel codimension-two symmetry defects with duality/triality defects were studied in \cite{Choi:2022zal}.} implementing the one-form symmetry transformation with parameter $n_x  = -{n \over N}$ due to $T_x t_x$ of (\ref{anomaly}).  The careful study of the various  defects mentioned should make finding these fusion rules possible, but we leave this interesting question to future studies. We  believe, however, that the $\T^4$ path-integral discussion after Eqn.~(\ref{twistedZ}) gives strong support for our finding  (\ref{anomaly})---while, admittedly, leaving the study of the more general interesting situations mentioned above open. 

\bigskip

  \bibliography{RefGaugedU1NonInv.bib}
 
  \bibliographystyle{JHEP}

  \end{document}